              \documentclass[final,1p,times]{elsarticle}

\usepackage{amssymb, bm,amsbsy}
\usepackage{amsthm, slashed}

\journal{Physics Letters B}

\begin{document}

\begin{frontmatter}



\title{Ferromagnetic phase diagram of neutron matter}

 \author[label1]{J.P.W. Diener\corref{cor1}}
 \ead{jpwd@sun.ac.za}
 \author[label1,label2]{F.G. Scholtz}
 \address[label1]{Insitute of Theoretical Physics, Stellenbosch University, P.O. Box X1, Matieland, 7602, South Africa}
 \address[label2]{National Insitute of Theoretical Physics, P.O. Box X1, Matieland, 7602, South Africa}
 \cortext[cor1]{Corresponding author}



\begin{abstract}
The magnetic properties of matter under extreme conditions are of particular importance to understanding the neutron star interior.  One contributing factor to the magnetic field of a neutron star could be the ferromagnetic phase of nuclear matter. In this letter we present a self-consistent, relativistic description of ferromagnetism in dense matter, within which the ferromagnetic phase diagram for neutron matter is calculated.
\end{abstract}


\end{frontmatter}


The properties of dense nuclear matter are of interest to various branches of topical research. One area of interest is the description of the constituent matter of neutron stars. Neutron stars are extreme laboratories: extremes in pressure, density and magnetic field are just some of the factors that need to be considered when studying neutron star matter. The neutron star interior is believed, at least in part, to consist of nuclear (baryonic) matter. Since charge neutrality of the star is required (a nett charge would result in a Coulomb force that will rip the star apart) the majority of the baryonic matter should be neutrons \cite{csg}.
 
The origin of the very strong magnetic field of neutron stars has been a topic of discussion for some time. Brownell and Callaway \cite{BN} as well as Silverstein \cite{S} proposed that the ferromagnetic phase of nuclear matter can make a significant contribution to the magnetic field of a neutron star.  This idea was investigated by various authors, using both non-relativistic and relativistic models, most recently Modarres and Pourmirjafari \cite{iran}, which also provides references to previous investigations. The outcome of these investigations seems to be inconclusive regarding the existence of a ferromagnetic phase of neutron matter and appear to be very model-dependent.

Due to the extreme densities encountered in the neutron star interior relativistic effects become more pronounced and thus it has been argued \cite{csg,walecka} that a relativistic description for the interior of the star may be necessary. Thus, to investigate ferromagnetic effects in the baryonic part of the neutron star interior a relativistic description of ferromagnetism in neutron and nuclear matter needs to be considered. Such a description of ferromagnetism was investigated by Maruyama and Tatsumi \cite{jp1}, which based their work on that of Niembro et al (see \cite{pr1} and references therein). We aim to further this work by presenting a self-consistent calculation of the magnetic field. We believe that a self-consistent approach is well-suited to this problem since the origin of the magnetic field in the ferromagnetic phase is the nucleon magnetic dipole moments, which in turn reacts to the presence of a magnetic field. 

We perform this self-consistent calculation in the relativistic mean-field approximation.  This is done by coupling the magnetic dipole moment of the neutron to the magnetic field of the ferromagnetic phase within the relativistic description pioneered by Walecka \cite{walecka} called Quantum Hadrodynamics (QHD). QHD is an effective field theory model for nuclear matter with mesons as degrees of freedom. It had undergone various extensions and modifications, as reviewed in \cite{serot} and has been used extensively to study the properties of nuclei, nuclear matter and neutron stars; see for instance \cite{csg,piek,meng,diener}. We employ the coupling between the dipole moment and the magnetic field which was introduced by Broderick et al \cite{broderick} to investigate the interaction between the magnetic dipole moment of nucleons and an external magnetic field. Within the relativistic mean-field approximation we use the Euler-Lagrange equation to derive an equation of motion for the ferromagnetic magnetic field (\ref{MFT1Bz}), which we then solve self-consistently.  As far as we can establish such a self-consistent calculation using this coupling has not been performed previously. 

To describe the interaction between the neutrons we couple the nucleon field $\psi$ to a scalar (sigma) $\phi$ and vector (omega) $\omega^\mu$ meson field and to the photon field tensor $F^{\mu\nu}$. The $\phi$ meson describes the long-range attraction of the NN-potential, while the $\omega$ mesons accounts for the short-range repulsion \cite{walecka}. This description is valid for cold, neutral neutron matter, since the neutron matter is considered to be at zero temperature and the included mesons have no charge. The Lagrangian is
\begin{eqnarray}
{\cal L} = 
{\cal L}_{\mbox{{\footnotesize Dirac}}}+
{\cal L}_{\mbox{{\footnotesize KG}}}+
{\cal L}_{\mbox{{\footnotesize Proca}}}+
{\cal L}_{\mbox{{\footnotesize EM}}}
-g_{v}\bar{\psi}\gamma_{\mu}\omega^{\mu}\psi
-g_b\bar{\psi}F^{\mu\nu}\sigma_{\mu\nu}\psi
+g_{s}\bar{\psi}\phi\psi.
\end{eqnarray}
Here ${\cal L}_{\mbox{{\footnotesize Dirac}}}$ is the free field component of the Lagrangian for the neutrons, ${\cal L}_{\mbox{{\footnotesize KG}}}$ is the Klein-Gordon Lagrangian for the scalar mesons and likewise ${\cal L}_{\mbox{{\footnotesize Proca}}}$ is the Proca Lagrangian for the massive vector mesons. ${\cal L}_{\mbox{{\footnotesize EM}}}$ is the free-field component of the electromagnetic Lagrangian. The coupling of the mesons to the different nucleon densities is the standard coupling as in \cite{walecka}. $\sigma_{\mu\nu}$ are the generators of the Lorentz group. If $A^0 = 0$ there is no electric field and $F^{\mu\nu}\sigma_{\mu\nu}$ reduces to $-2B_i\Sigma_i$ in a co-moving frame of reference, with $B_i$ the components of the magnetic field and $\Sigma_i$ the three spatial four-component spin matrices.

For the coupling strengths $g_v$ and $g_s$ the values of the QHD1 parameter set given in \cite{walecka} were used. These coupling constants are chosen to fit various properties of nuclear matter at the nuclear saturation point ( $0.16 fm^{-3}$).  The strength of the nucleon and electromagnetic coupling $g_b$ is left as a free parameter with the units of the nuclear magnetic dipole moment ($C fm$).

The magnetic field is solved for in a co-moving frame of reference chosen, without loss of generality, so that the magnetic field lies in the $z$-direction. 

In the relativistic mean-field (RMF) approximation the nucleon and meson operators are replaced by their groundstate expectation values \cite{walecka}. The equations of motion of the various fields are then found to be
\begin{eqnarray}
		\phi &=& \frac{g_{s}}{m_{\sigma}^{2}}\left\langle \bar\psi\psi\right\rangle\label{MFT1sigma} \\
		\omega_{0} &=& \frac{g_{v}}{m_\omega^{2}}\left\langle\psi^\dagger\psi\right\rangle\label{MFTomega}\\
		B_z &=& -\frac{g_b}{e^2}\left\langle \bar\psi\Sigma_z\psi\right\rangle\label{MFT1Bz}\\
 		0&=&\Big[i\slashed{\partial}-g_{v}\gamma_0\omega_{0} -g_b B_z\Sigma_z - 
		\big(m-g_{s}\phi\big)\Big]\psi\label{MFT1nucleon}.
	\end{eqnarray}
For the equation of motion of the $\omega$ meson field in (\ref{MFTomega}), the RMF approximation results in the spatial components being zero and thus only the zeroth component is considered \cite{serot}.

From (\ref{MFT1nucleon}) it can be seen that the equation of motion of the nucleons is a modified version of the Dirac equation:
\begin{eqnarray}
		\big[i\partial_t-g_{v}\omega_{0}\big]\psi=\big[-i{\bm \alpha}\cdot\nabla  +g_bB_z\beta\,\Sigma_z + 
		\beta m^*\big]\psi,\label{dirac}
	\end{eqnarray}
with $m^* = m-g_{s}\phi$ the effective mass and $\bm \alpha$ and $\beta$ the Dirac matrices.

Thus the nucleon (Dirac) field $\psi$ can be constructed in analogy to the free particle solutions of the Dirac equation \cite{walecka}. Therefore $\psi$ is assumed to have the form
\begin{equation}
	\psi_{\bm k,\lambda}(x) \propto \psi({\bm k},\lambda)\,e^{i{\bm k}\cdot {\bm x} - ie({\bm k},\lambda)t}\label{Kexpand}
\end{equation}
with $e({\bm k},\lambda)$ the single particle energy. Since a non-zero magnetic field would break the spherical symmetry of the ground state it is convenient to express $\bm{k}$ in cylindrical coordinates as
\begin{eqnarray}
	{\bm k} = (k_\bot, k_z) =\big(\sqrt{k_x^2+k_y^2},\ k_z\big).
\end{eqnarray}
As pointed out by previous authors, \cite{jp1} and references therein, care has to be taken when considering the magnetic dipole moment (spin) of nucleons in the context of the Dirac equation. This is due to the fact that the spin operator does not commute with the Dirac Hamiltonian, the matrix on the right hand side of (\ref{dirac}), and therefore spin is not a good quantum number.  Using the form of the wavefunction in (\ref{Kexpand}), the single particle energies are found to be
\begin{eqnarray}
	e({\bm k},\lambda)-g_v\omega_0 	= 
	\pm\sqrt{\left(\sqrt{k_{\bot}^2+{m^*}^2} +\lambda g_b B_{z}\right)^2+k_{z}^2}\label{singlepatE}.
\end{eqnarray}
Here $e({\bm k},\lambda)$ is labeled by $\lambda = \pm 1$, which refers to the contribution of the magnetic field to the single particle energies. $\lambda$ will serve to differentiate between the two species of neutrons, but this should not be mistaken for a spin label.  As in the case of the unmodified Dirac equation both positive (particle) and negative (antiparticle) energies are found. 

Using the notation of (\ref{Kexpand}) the unnormalised Dirac spinors for particle are
\begin{eqnarray}
							\psi(\bm k, \lambda) = \left[\begin{tabular}{c}
								$\frac{k_\bot^2+[e({\bm k},\lambda)+g_b B_z+m^*]
								[m^*+\lambda\sqrt{k_\bot^2+{m^*}^2}]}
								{[k_x+ik_y][e({\bm k},\lambda)+g_b B_z+\lambda\sqrt{k_\bot^2+{m^*}^2}]}$\\
								\\
								$\frac{-k_z}{e({\bm k},\lambda)+g_b B_z+\lambda\sqrt{k_\bot^2+{m^*}^2}}$\\
								\\
								$\frac{k_z[m^*+\lambda\sqrt{k_\bot^2+{m^*}^2}]}
								{[k_x+ik_y][e({\bm k},\lambda)+g_b B_z+\lambda\sqrt{k_\bot^2+{m^*}^2}]}$\\
								\\
								$1$\\
							\end{tabular}
						\right].\label{spinor}
\end{eqnarray}
The spinors are normalised so that
\begin{eqnarray}
	\int d^{3}\!x\ \psi^\dagger_{\bm k',\lambda'}(\!x)\ \psi_{\bm k, \lambda}(\!x) = \delta(\bm k'-\bm k)\delta_{\lambda,\lambda'}.
\end{eqnarray}

\begin{figure}
	\centering
		\includegraphics[width=0.7\textwidth]{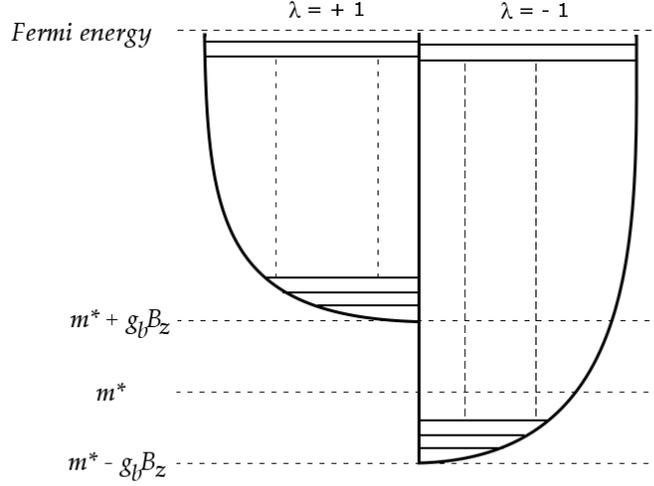}
	\caption{Illustration of the single particle energy levels that neutrons can populated in the presence of a magnetic field. The unpaired neutrons are of course the source of the magnetic field.}
	\label{fig:fermi}
\end{figure}

By inspection of the single particle energies (\ref{singlepatE}) it clear that when the magnetic field is zero, the normal dispersion relation is recovered and the lowest single particle energy state that a fermion can occupy is the effective mass, $m^*$. In the presence of a magnetic field the lowest single particle energy would be $m^* \pm g_b B_z$ for $\lambda = \pm 1$. The magnetic field therefore lifts the degeneracy between the states of the two neutron species. This splitting is illustrated in Fig.$\:$(\ref{fig:fermi}) for positive energy states, but also applies to the negative energy states. However the groundstate we consider only has positive energy nucleon states, which is filled to a certain energy, the Fermi energy.

Using (\ref{Kexpand}) to calculate the nucleon densities on the right hand side of equations (\ref{MFT1sigma} - \ref{MFT1Bz}), it can be shown that these equations are equivalent to the minimization of the energy density of the system, $\epsilon$,
\begin{eqnarray}
\epsilon = 
\sum_\lambda\int\frac{d{\bm k}}{(2\pi)^3}\,e({\bm k},\lambda)\,\Theta\big[\,\mu-e({\bm k},\lambda)\big] 
+\frac{1}{2}m_{\sigma}^{2}\phi^{2}+ \frac{1}{2}m_\omega^{2}\omega_{0}^2
		 +\frac{1}{2}e^2B_z^2\,,
\end{eqnarray}
with $\mu$ the chemical potential and $\Theta$ a step function to ensure that only energies below the Fermi energy are considered. 

\begin{figure}
	\centering
		\includegraphics[width=0.7\textwidth]{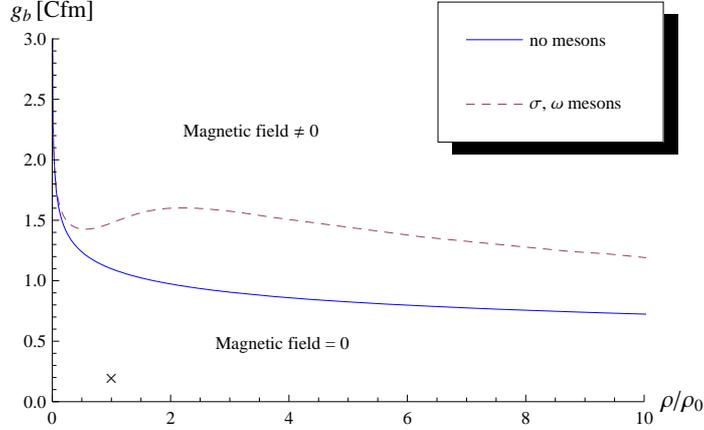}
	\caption{Ferromagnetic phase diagram in density and coupling constant, $g_b$. The solid line represents the phase boundary when no mesons are included in the description and thus the neutrons only interact with the magnetic field. The dotted line represents the phase boundary when mesons are also included. The $\times$ indicates the point where $g_b$ is at the value of the magnetic dipole moment of the neutron at a density which is equal to nuclear saturation density $\rho_0 = 0.16\ fm^{-3}$ .}
	\label{fig:gb}
\end{figure}

We used the equations of motion to calculate the values of the different fields at a given density and value of the coupling $g_b$. The coupling constant was increased at a fixed density until a non-zero value of the magnetic field was found. The results are shown in Fig.$\:$(\ref{fig:gb}). In this figure the solid line represents the coupling strength $g_b$ at a specific density which would effect a ferromagnetic phase transition when no mesons are present. The dashed line represents the case when mesons are included. 

We note that the vector mesons do not influence the phase transition since they couple to the nucleon density and merely shift the energy spectrum. This can be seen in the expression for the single particle energies (\ref{singlepatE}). 

The effect of the scalar meson is to reduce the mass of nucleon resulting in an effective mass $m^*$. This reduction in the mass implies that for the same density the coupling constant $g_b$ has to be stronger than in the case when no mesons are present. Fig.$\:$(\ref{fig:gb}) clearly shows that the inclusion of the scalar mesons result in the phase transition occurring consistently at larger values of $g_b$.

The shape of the curve in Fig.$\:$(\ref{fig:gb}) at low densities and with mesons included, results from the interplay between the contributions from the scalar and nucleon fields to the energy density. 

From Fig.$\:$(\ref{fig:gb}) it is apparent that for the ferromagnetic phase to present itself in cold neutron matter the coupling between the nucleon field and the magnetic field has to have some density dependence. The coupling at normal nuclear density (the magnetic dipole moment of the neutron) is just not strong enough to affect a phase transition even at very high densities. The density dependence of $g_b$ might be calculated by including renormalization effects, but this was not investigated in the current work. 

It is possible that the densities at which the ferromagnetic phase of neutron matter is stable can be reached in neutron stars. In the calculation done here such a stable phase generates a magnetic field of the order of $10^{10}$ Gauss. This magnitude of the field in itself is not enough to explain the origin of the magnetic field of neutron stars, which are calculated to have surface magnetic fields of the order of $10^{11}\!-10^{13}\:G$ \cite{harding}. The model presented here is a first approximation for neutron star matter, since only neutrons are included and only infinite matter considered. Further work would entail studying the effects of including other hadrons and leptons in the model to obtain a more realistic description for neutron star matter. With such a model it might be possible to investigate the effects of a ferromagnetic phase on the equation of state of neutron star matter. 

\section{Acknowledgements}
This research is supported by the South African SKA project and a grant from the National Research Foundation of South Africa.


\begin{thebibliography}{00}
 \bibitem{csg} N. K. Glendenning, Compact Stars, Nuclear Physics, Particle Physics and General Relativity, 2\raisebox{1.5mm}{\tiny{nd}} edition, Springer, New York 2000.
 \bibitem{BN} D.H. Brownell, J. Callaway, Nuovo Cimento 60B (1969) 169.
 \bibitem{S} S.D. Silverstein, Physical Review Letters 23 (1969) 139. 
 \bibitem{iran} M. Modarres, T. Pourmirjafari, Nuclear Physics A 836 (2010) 91.
 \bibitem{walecka} B.D. Serot, J.D. Walecka, The Relativistic Nuclear Many-Body Problem, in: J.W. Negele, E. Vogt (Eds.) Advances in Nuclear Physics 16 (1986) 1.
 \bibitem{jp1} T. Maruyama, T. Tatsumi, Nuclear Physics A 693 (2001) 710.
 \bibitem{pr1} R. Niembro, S. Marcos, M.L. Quelle and J. Navarro, Physics Letters B 249 (1990) 373.
 \bibitem{serot} B.D. Serot, J.D. Walecka, International Journal of Modern Physics E 6 (1997) 515.
 \bibitem{piek} C.J. Horowitz, J. Piekarewicz, Physical Review Letters 86 (2001) 5647.
 \bibitem{meng} J. Meng, H. Toki, S.G. Zhou, S.Q. Zhang, W.H. Long and L.S. Geng, Progress in Particle and Nuclear Physics 57 (2000) 470.
 	\bibitem{diener} J.P.W. Diener, Relativistic mean-field applied to the study of neutron star properties, MSc thesis, Stellenbosch University 2008, ArXiv:0806.0747.
 	\bibitem{broderick} A. Broderick, M. Prakash and J.M. Lattimer, The Astrophysical Journal 537 (2000) 351.
	\bibitem{harding} A. Harding, D. Lai, Reports on Progress in Physics 69 (2006) 2631.
\end{thebibliography}
\end{document}